\documentclass[lettersize,journal]{IEEEtran}
\usepackage{amsmath,amsfonts}
\usepackage{algorithmic}
\usepackage{algorithm}
\usepackage{array}
\usepackage[caption=false,font=normalsize,labelfont=sf,textfont=sf]{subfig}
\usepackage{textcomp}
\usepackage{stfloats}
\usepackage{url}
\usepackage{verbatim}
\usepackage{graphicx}
\usepackage[numbers]{natbib}

\newcolumntype{L}[1]{>{\raggedright\let\newline\\\arraybackslash\hspace{0pt}}m{#1}}
\newcolumntype{C}[1]{>{\centering\let\newline\\\arraybackslash\hspace{0pt}}m{#1}}
\newcolumntype{R}[1]{>{\raggedleft\let\newline\\\arraybackslash\hspace{0pt}}m{#1}}

\begin{document}

\title{Test case quality: an empirical study on belief and evidence}

\author{{Daniel Lucrédio, Auri Marcelo Rizzo Vincenzi, Eduardo Santana de Almeida, Iftekhar Ahmed}
\thanks{Daniel Lucrédio and Auri Marcelo Rizzo Vincenzi are with the Computing Department, Federal University of São Carlos, Rod. Washington Luís, Km~235, P.O. Box 676 - 13565-905, São Carlos - SP - Brazil (e-mail: daniel.lucredio@ufscar.br; auri@ufscar.br)

Eduardo Santana de Almeida is with the Computing Institute, Federal University of Bahia, Avenida Milton Santos, s/n - Campus de Ondina, PAF 2, CEP: 40.170-110 Salvador-Bahia (e-mail: esa@rise.com.br) 

Iftekhar Ahmed is with the Donald Bren School of Information and Computer Science at the University of California, Irvine, 2438 ISEB, Irvine, CA 92697-3440 (e-mail: iftekha@uci.edu)}
}



\maketitle

\begin{abstract}
Software testing is a mandatory activity in any serious software development process, as bugs are a reality in software development. This raises the question of quality: good tests are effective in finding bugs, but until a test case actually finds a bug, its effectiveness remains unknown. Therefore, determining what constitutes a good or bad test is necessary. This is not a simple task, and there are a number of studies that identify different characteristics of a good test case. A previous study evaluated 29 hypotheses regarding what constitutes a good test case, but the findings are based on developers' beliefs, which are subjective and biased. In this paper we investigate eight of these hypotheses, through an extensive empirical study based on open software repositories. Despite our best efforts, we were unable to find evidence that supports these beliefs. This indicates that, although these hypotheses represent good software engineering advice, they do not necessarily mean that they are enough to provide the desired outcome of good testing code.
\end{abstract}



\section{Introduction}
\label{sec:intro}
Testing is an indispensable part of the software development process. The goal of software testing is to make sure that a software satisfies its functional and non-functional requirements as mentioned in the requirements document. Past studies have shown the importance of testing and the significant impact it can have on the economy~\citep{Tassey2002}.

Even with the significant amount of time invested in testing \citep{kunar:procedia:2016}, bugs are common in software \citep{Habib:2018:ase,Adlemo:2018:wret,Aniche:2021:arxiv}. This brings forward the question of test quality and the characteristics exhibited by good test cases. A poor quality test case will fail to serve its purpose in the first defensive line in bug prevention \citep{Grano:2019:issta}. It may also have a strong negative impact on program comprehension and maintenance \citep{bavota:2012:icsm,bavota:2014:ese}. Therefore, being able to correctly determine the quality of a test case is very important for software organizations \citep{bavota:2014:ese}.

Determining what constitutes a good or bad test case is not simple \citep{Bowes:2017:wetsm}. The main desired property is effectiveness, i.e. the ability to reveal bugs \citep{Tosun:2018:icssp,Grano:2019:issta}. Organizations try to estimate effectiveness based on metrics such as code coverage \citep{BRUNTINK20061219,Edwards:2014:icse} or mutation score \citep{Tosun:2018:icssp}. While effectiveness is certainly important, there are other characteristics that should be exhibited by good test cases \citep{Bowes:2017:wetsm}. Principles such as simplicity \citep{Bowes:2017:wetsm}, ease of understanding and maintenance \citep{Grano:2019:issta}, and even more elusive goals, such as to minimize the safety-related lawsuit risk \citep{Adlemo:2018:wret}, should also guide the development of test cases. In addition to the great variety of characteristics that must be considered, there is also the issue of subjectivity. Each team or organization may have different views on what is important and what is not.



In an attempt to better understand these qualitative aspects, a previous study \citep{pavneet2019icse} interviewed 21 and surveyed 261 practitioners to understand what makes good test cases and their views on software testing. The study identified 29 hypotheses using the interviews that were validated in the survey spanning multiple dimensions such as test case contents, size and complexity, coverage, maintainability, and bug detection. For these hypotheses, the study listed reasons on why practitioners agree or disagree and highlighted open problems faced by practitioners.

The above study was largely qualitative, based on the opinions of software practitioners, which raises a question: are these opinions based on facts or are unjustified beliefs expressed by practitioners? Devanbu et al. studied the beliefs of software practitioners and the relationship between these beliefs and the data from the projects these practitioners work on~\citep{devanbu2016belief}. Their findings show that practitioners have very strong beliefs that are based on personal experiences rather than empirical research and their beliefs can be different from the evidence of the project. The study further suggests that ``more in-depth study the interplay of belief and evidence in software practice is needed''.

Following Devanbu et al.'s suggestion, in this paper we report the results of an empirical validation of eight of the hypotheses formulated in the aforementioned study \citep{pavneet2019icse}. We tried to confirm these hypotheses by looking into 42 popular repositories, in an attempt to find evidence to support the practitioners' opinions. Our main metric is the number of bug fixing commits associated with a test case in the software's history. Differently from mutation score, which is an estimate \citep{Grano:2021:ieeetse}, this serves as an objective measurement of how effective a test case was in finding actual bugs. We analyzed the correlation between this metric and other dynamic and static metrics that capture the essence of the different hypotheses being evaluated. 

The observed repositories are of various sizes and come from popular organizations such as Google, Microsoft, Twitter and Netflix, having a large number of active contributors. Therefore we considered these as examples of good software projects that follow solid software engineering principles. 

Despite our best efforts, we were able to find weak evidence supporting only one of the hypotheses. This observation is important because it suggests that these beliefs are not necessarily met by evidence found in popular software repositories. This could mean that they are based on personal belief rather than observable outcomes of good testing code, as suggested by \citet{devanbu2016belief}.



The remainder of this paper is structured as follows. Section \ref{sec:related} discusses related work. Section \ref{sec:method} describes
our methodology. Section \ref{sec:result} presents the results. We discuss
implications and threats to validity in Sections \ref{sec:implications} and \ref{sec:threats}. We conclude and present future work in Section \ref{sec:conclusion}.

\section{Related Work}
\label{sec:related}


The best way to ensure software quality is to do extensive and thorough software testing with the goal of discovering faults in the system by executing test cases. 
As testing is one of the most expensive part of software development~\citep{kracht2014empirically}, it is important to measure its quality. The quality of a test suite is measured by the number of real faults it can detect in the code~\citep{gligoric2013comparing}. Researchers proposed different measures of test suite quality, mostly focused on code coverage~\citep{gligoric2013comparing,gopinath2014code,groce2014coverage}. Researchers also use mutation analysis~\citep{demillo1978hints,hamlet1977testing,jia2010analysis} to measure the test quality, seeding simple syntactic changes to the program and measuring the ability of the test suite to distinguish the semantic difference.

Others have investigated more abstract quality attributes of a test case. \citet{Adlemo:2018:wret} have compiled 15 quality criteria based on previous studies and discussions with industry partners. They analyzed these criteria by sending a questionnaire to randomly chosen respondents in the Swedish software industry, but only 13 provided answers. The results are not very conclusive.


Researchers proposed different ways to improve testing. \citet{van2007detection} used the concept of unit testing to propose a set of metrics that detect test smells -- a concept similar to code smells. They suggested that a reliable test smells detection tool is necessary to improve test quality. Another study, by \citet{greiler2013automated}, investigated industrial projects and found that test smells can also be caused by fixture set-up. They also provide a tool, TestHound, that uses static analysis technique to identify fixture-related test smells and provide recommendations for refactoring the smelly test code. \citet{palomba2017does} investigated the relationship between flaky tests and test smells (more specifically Resource Optimism, Indirect Testing and Test Run War test smells). They studied 18 software systems and observed that 54\% of tests that are flaky contain a test code smell that can cause the flakiness.


Most of the studies described above on software testing investigate only the artifacts (for example, code and bug reports) rather than the opinion of actual practitioners. However, in a case study at Microsoft, \citet{devanbu2016belief} found that there is a difference between empirical
evidence (quantitative evidence) and practitioner’s belief from personal experiences in software engineering settings. Researchers have conducted different studies to investigate practitioners’ perception of various software engineering research work~\citep{begel2014analyze,lo2015practitioners,meyer2014software,zou2018practitioners,kochhar2016practitioners}. \citet{cartaxo2016evidence} conducted a survey and found that practitioners rarely use systematic review research papers to acquire knowledge. \citet{lo2015practitioners} conducted a survey of 512 practitioners from Microsoft to rate software engineering research relevant to the practitioners in the field and found that 
practitioners were generally positive about the studies done by the research community. They also revealed
several reasons why practitioners considered certain research ideas to be unwise. \citet{begel2014analyze} conducted a survey among software engineers and summarized 145 questions that software engineers mostly wanted data scientists to solve.

Besides the above studies, researchers have also investigated the differences between empirical evidence and practitioners’ perceptions on some specific software engineering techniques~\citep{palomba2014they,kochhar2016practitioners,zou2018practitioners}. \citet{kochhar2016practitioners} investigated
how practitioners appreciate the  existing research work on automated
fault localization techniques by surveying 386 practitioners from 30 different countries. In this study the authors investigate the factors that impact practitioners' willingness to adopt a fault localization technique and compared practitioners need with state-of-research. \citet{zou2018practitioners} conducted  a survey among 327 practitioners to understand whether the automated bug report management techniques from research community is actually required and applicable for practitioners. They asked the survey respondents to rate the automated bug report management techniques and provide some potential research directions in developing techniques to help developers better manage bug reports. \citet{palomba2014they} studied how developers treat bad code smells detected by the research community. More specifically, they investigated the extent developers considered code smells as design problems and which code smells developers considered as threats and found that characteristics of code smells are not yet explored sufficiently by the developers. \citet{daka2014survey} and \citet{kochhar2016practitioners} investigated practitioners’ views on testing practices.  


\citet{daka2014survey} conducted a survey to better understand the practice of unit testing among developers. They investigated aspects such as motivation, usage of automation tools and their challenges. \citet{kochhar2016practitioners} investigated the practitioners’ views on what make good test cases. In this study, the authors interviewed 21 practitioners and created a list of 29 hypotheses that describe characteristics of good test cases and testing practices and then validate those hypotheses by conducting a survey among 261 practitioners. None of these two studies investigate the empirical evidence for the practitioners' views. In this work, we attempt to find the empirical evidence for the hypotheses investigated by \citet{kochhar2016practitioners}, so that we can find if there is any differences in between practitioners' perspectives and empirical evidence. Doing this we can highlight the areas that need more attention by the research community.

\citet{Grano:2021:ieeetse} studied the relation between 67 factors related to both production and test code and test case effectiveness, measured by means of mutation score. Their study confirmed that there are medium to strong correlations between 41 of these factors and mutation score. These factors include static metrics such as Lines Of Code, Lack of Cohesion and Cyclomatic Complexity. Together, they constitute a lightweight approach to estimate mutation score with good accuracy (around 86\% of F-Measure and AUC-ROC). But as highlighted by the authors, mutation score is only an approximation of test effectiveness. This is where our study differs from \citet{Grano:2021:ieeetse}'s. We studied less factors, but we tried to measure the actual test effectiveness by counting the number of actual bug fixes associated with a test case in the software's history.
\section{Methodology}
\label{sec:method}

This study attempted to find empirical evidence for eight of the hypotheses investigated in a previous research \citep{pavneet2019icse}, by looking into the information contained in public software repositories, which consists mainly of source code and versioning information, including commit messages, tags and other metadata. 

From the 29 hypotheses of the previous research (First three columns of Table \ref{tab:hypotheses}), we initially tried to identify which ones to investigate. Because our investigation used data available from source code repositories, we had to exclude from the analysis those hypotheses that require information not available, such as domain-specific and subjective information. In this sense, we tried to classify each hypothesis in terms of how difficult it would be to collect empirical evidence from the repositories (Fourth column of Table \ref{tab:hypotheses}). 24 hypotheses were classified as ``possible'' (i.e. the information can be extracted), 3 were classified as ``impossible'' (i.e. the information can not be extracted). 2 hypotheses were discarded because they have already been proved elsewhere: \citet{kochhar2015code} and \citet{Barani2023} have addressed H11, demonstrating the existence of other factors, in addition to code coverage, that are important for test effectiveness. Regarding H27 (determinism in test cases), there is a large consensus that confirms this hypothesis, both from industry and academia, as reported by \citet{Ziftci2020}. \citet{Rahman2018} also demonstrated in experiments that non-determinism hinders a test case's ability to prevent crashes.

In summary, for this study we ended up investigating eight hypotheses (Last column of Table \ref{tab:hypotheses}), from groups ``Contents'', ``Size and complexity'' and ``Coverage''. Others are left for future work.

\begin{table*}[htb!]
    \centering
    \caption{List of hypotheses from \citep{pavneet2019icse}. The third column corresponds to the average Likert score from the survey response. The fourth column indicates the considered difficulty for the hypothesis: P=Possible, I=Impossible, AP=Already proved. The fifth column indicates whether the hypothesis was included here (Y) or not (N).}
    \label{tab:hypotheses}
        \begin{tabular}{|c|p{8.2cm}|c|c|c|} \hline

        \multicolumn{5}{|c|}{\textbf{Contents}} \\ \hline
        H1 & A good test case is specific or atomic, i.e., one test case should be testing one aspect of a requirement. & 3.93 & P & Y \\
        H2 & Test cases in a test suite should be self-contained, i.e., independent of one another. & 3.95 & P & Y \\
        H3 & Good test cases should check for normal and exceptional flow. & 4.47 & P & Y \\
        H4 & Test cases must perform boundary value analysis, i.e., take as input values at the extreme ends of an input domain. & 4.24 & P & N \\
        H5 & Test cases should serve as a good reference documentation. & 3.93 & P & N \\ \hline

        \multicolumn{5}{|c|}{\textbf{Size and complexity}} \\ \hline
        H6 & Most test cases should be small in size (in terms of its lines of code). & 3.85 & P & Y \\
        H7 & Large test cases are often hard to understand and maintain. & 3.73 & P & Y \\
        H8 & Large test cases may be needed to detect difficult bugs. & 3.59 & P & N \\
        H9 & A good suite contains lots of small test cases (with fewer LOC) and few large test cases. & 3.97 & P & N \\
        H10 & Increased complexity in a test case can lead to bugs in the test code itself. & 4.04 & P & Y \\ \hline
    
        \multicolumn{5}{|c|}{\textbf{Coverage}} \\ \hline
        H11 & Code coverage is necessary but not sufficient. & 3.97 & AP & N \\
        H12 & Code coverage should be used to understand what is missing in the tests and create tests based on that. & 3.97 & P & N \\
        H13 & Higher coverage does not mean that a test suite can detect more bugs. & 4.02 & P & N \\
        H14 & Each test case should have a small footprint, i.e., the amount of code it executes. & 3.92 & P & Y \\
        H15 & A test case that is designed to maximize coverage is often long, not understandable and brittle (i.e., breaks easily). & 3.50 & I & N \\
        H16 & Designing test cases to cover different requirements is often more important than designing test cases to cover more code. & 4.00 & P & N \\ \hline

        \multicolumn{5}{|c|}{\textbf{Maintainability}} \\ \hline
        H17 & A good test case should be well-modularized. & 4.62 & P & N \\
        H18 & A good test case should be readable and understandable. & 4.58 & P & Y \\
        H19 & Test cases should be simpler than the code being tested. & 4.20 & P & N \\
        H20 & Test code should be designed with maintainability in mind since evolution of code often requires changing of test code. & 4.16 & P & N  \\
        H21 & Traceability links should be maintainable between test code, requirements, and source code. & 3.97 & P & N \\ \hline

        \multicolumn{5}{|c|}{\textbf{Bug detection}} \\ \hline
        H22 & A good test case should attempt to break functionality to find potential bugs. & 4.11 & I & N \\
        H23 & Test even the simplest things that cannot go wrong. & 3.89 & I & N \\
        H24 & During maintenance, when a bug is fixed, it is good to add a test case that covers it. & 4.40 & P & N \\
        H25 & Test assertions can help detect subtle errors that might otherwise go undetected. & 4.51 & P & N \\
        H26 & Adding common errors and possible causes as comments in test code is helpful to debug failures. & 3.98 & P & N \\ \hline
        
        \multicolumn{5}{|c|}{\textbf{Others}} \\ \hline
        H27 & A good test case should be designed such that its results are deterministic. & 4.07 & AP & N \\
        H28 & Test cases in a test suite should not have side effects so running a test before or after another should not change the results. & 4.28 & P & N \\
        H29 & Test cases should use tags or categories, such as slow tests, fast tests etc., so as to be able to run a specific set of tests easily at a time. & 3.93 & P & N \\ \hline

    \end{tabular}

\end{table*}


We decided that the dataset should be composed of projects that are known to have good quality in terms of code standards and good software engineering practices, because this is important for some of the metrics, as discussed later. In order to achieve such goals, we only selected open source projects that are sponsored by well-known organizations or individuals and have high maturity. This was a manual process. We started by looking at repositories from Google, Microsoft, Twitter, Netflix and Amazon. Later we included Hibernate, Apache and Spring, which are large software projects for Java. Next we included other repositories, trying to diversify the dataset, but no systematic approach was followed other than looking for long-lived repositories (at least five years into development) with an active community (commits in the last six months).

Three additional constraints were used to select the projects:

\begin{itemize}
\item The project had to be on github.com, to allow automatic retrieval of commit information;
\item The project had to use Java with Maven or Gradle, to allow automatic dependency resolution, static code analysis and coverage analysis;
\item Our code analysis scripts rely on code annotations, therefore we had to restrict the projects to a known framework. We chose JUnit versions 4 or 5, AssertJ\footnote{\url{https://assertj.github.io/doc/}}, Truth\footnote{\url{https://github.com/google/truth}} and Hamcrest\footnote{\url{https://hamcrest.org/}}.
\end{itemize}

In the end, a total of 42 projects, listed in Table \ref{tab:projects}, were selected.

\begin{table}[htb]
    \caption{Projects used in this research}
    \label{tab:projects}
        \centering
    \begin{tabular}{|c|C{5.5cm}|} \hline

\textbf{\shortstack[c]{Repository name \\ (in github.com)}} & \textbf{Projects} \\ \hline

google

& guice, gson, closure-compiler, open-location-code, google-java-format, error-prone, truth, jimfs, compile-testing, closure-templates \\ \hline

microsoft

& spring-cloud-azure, spring-data-cosmosdb, spring-data-gremlin, Git-Credential-Manager-for-Mac-and-Linux \\ \hline

twitter

& Nodes, hraven, ambrose, GraphJet, hbc \\ \hline

Netflix

& zuul, mantis, hollow, governator, Fenzo \\ \hline

hibernate

& hibernate-ogm, hibernate-search \\ \hline

amzn

& ion-java \\ \hline

apache

& incubator-datasketches-java \\ \hline

spring-io

& initializr \\ \hline

javaparser

& javaparser \\ \hline

plutext

& docx4j \\ \hline

jbehave

& jbehave-core \\ \hline

shopizer-ecommerce

& shopizer \\ \hline

kiegroup

& optaplanner-core \\ \hline

swagger-api

& swagger-core \\ \hline

rampatra

& jbot \\ \hline

NanoHttpd

& nanohttpd \\ \hline

digitalfondue

& lavagna \\ \hline

protostuff

& protostuff-core \\ \hline

easymock

& objenesis \\ \hline

oskopek

& javaanpr \\ \hline

greenmail-mail-test

& greenmail \\ \hline

    \end{tabular}

\end{table}

\subsection{Metrics}
\label{subsec:metrics}

To analyze the hypotheses, we used the following metrics, some of which already existed:

\begin{itemize}
    \item $loc(T)$: number of lines of code in test case $T$ (hypotheses H6 and H7);
    \item $cycl(T)$: McCabe's cyclomatic complexity of test case $T$ (hypotheses H7 and H10);
    \item $read(T)$: Posnett et al.'s software readability model for test case $T$, which is based on code size and entropy  \citep{posnett2011simpler} (hypothesis H18);
\end{itemize}

The following metrics were defined for this research:

\begin{itemize}
    \item $bfCommits(T)$: number of bug-fixing commits in the history of a test case (all hypotheses, except H7);
    \item $assert(T)$: number of assertions in a test case (hypothesis H1);
    \item $eCoupling(T)$: efferent coupling at method level (hypothesis H2);
    \item $eCouplingTC(T)$: efferent coupling for a test case (hypothesis H2);
    \item $invWithEx(T)$: number of method invocations with exceptions (hypothesis H3);
    \item $invWithExC(T)$: number of method invocations with exceptions thrown and caught (hypothesis H3);
    \item $footprint(T)$ - test case footprint (hypothesis H14);
    
\end{itemize}

Next we describe the metrics defined for this research.

\subsubsection{$bfCommits(T)$ - Number of bug-fixing commits in the history of a test case}

A software repository works with ``commits'', which is a collection of modifications to one or more source files made by a developer and submitted to the repository. A good software engineering practice (which we assume to be present in the projects in our dataset) is to ``commit often'', so that each commit has a single, well-defined purpose, such as introducing a new feature or fixing a bug. In our study, we are interested in bug-fixing commits. This metric counts, for each test case, the number of commits that involved at least one modification to that test case and that have the purpose of fixing a bug.


The rationale behind this metric is the following: in order to be considered ``good'', a test case has to fulfill its primary purpose of helping to find bugs. A well-known good practice in bug fixing is to write a new test case or modify an existing one in order to identify the bug. This may be done before or after the bug fixing itself, but the commit normally includes both changes: the new/modified test case and the bug-fixing changes. Therefore, in a bug-fixing commit, changes in a test case (creation or modification) will normally result in the correct identification of a bug. As a result, the more bug-fixing commits associated with a test case, the better it is.

This condition is used, explicitly or implicitly, in almost all hypotheses we tested (H1, H2, H3, H6, H10, H14 and H18).

This metric is defined as follows:

Let $P$ be the project where test case $T$ is defined. Let $H(P) = C_1, C_2, ..., C_n$ be the history of commits in the entire project $P$'s life cycle in the repository, $H_{tc}(P,T)$ be a subsequence of $H(P)$ containing only those commits that involved one or more modification to test case $T$, and $H_{tcbf}(P,T)$ be a subsequence of $H_{tc}(P,T)$ containing only those commits that have the single, well-defined purpose of fixing a bug, then: $bfCommits(T)$ is the number of elements in $H_{tcbf}(P,T)$.

\subsubsection{$assert(T)$ - Number of assertions in a test case}

\citet{xuan2014purification} have defined the concept of ``test case atomization'', which is the process where a single test case with $k$ assertions is replaced by $k$ single-assertion test cases. According to them, having a single assertion means that the test case can only fail for one reason, and this is why it is considered as an atomic test case. This metric aims at determining a test case's ``atomicity'' by counting the number of assertions in it, and is required for testing hypothesis H1.

When counting assertions, it is not enough to look into the code of the test case itself. Sometimes, the test case uses a helper method, and the assertion is in that helper method \citep{delplanque2019rotten}.

Following this rationale, $assert(T)$ is the number of assertions in test case $T$ and in the methods invoked by $T$, directly or indirectly. This metric is essentially the same as \citet{BRUNTINK20061219}'s dNOTC metric, although here we consider other frameworks in addition to JUnit, as mentioned before.

\subsubsection{$eCoupling(T)$ - Efferent coupling at method level}

This metric attempts to measure coupling, which is a well-known code modularity metric, but for a test case. A test case is a method, therefore we followed the same approach adopted by JArchitect\footnote{JArchitect is one of the most used tools for collecting Object-Oriented metrics. Its method-level metrics are described in \url{https://www.jarchitect.com/metrics#MetricsOnMethods}}, called efferent coupling at method level, which measures the number of methods a method directly depends on. This metric is required for testing hypothesis H2.

In this sense, this metric is defined as follows: let $S(T) = S_1, S_2, ..., S_n$ be the sequence of statements contained in test case T's body, and $I(T)$ be a subsequence of $S(T)$ containing only those statements that are method invocations, and $DM_{I}(T)$ be a set containing the distinct methods being invoked by statements from $I(T)$, then: $EfferentCoupling(T)$ is the number of elements in $DM_{I}(T)$.

\subsubsection{$eCouplingTC(T)$ - Efferent coupling for a test case}

This is essentially the same metric as the previous one, with one difference: ultimately, invoking methods from other classes is the very purpose of a test case, i.e. to exercise the system under test. Therefore, only method invocations that do not belong to the system under test should be considered as an indicative of coupling. However, it is a difficult task to determine which classes belong to the system under test and which do not, therefore we propose an approximation: we only include in this metric those invocations to methods that belong to the same class as the test case, because this is how dependency will probably be implemented by developers most of the cases. And because putting test cases in a different class than the methods being tested is a good software engineering practice, this approach excludes dependency from methods that belong to the system under test, which is what we wanted in the first place. This metric is also required for testing hypothesis H2.

Following this rationale, this metric is defined as follows: let $S(T) = S_1, S_2, ..., S_n$ be the sequence of statements contained in test case T's body, and $I(T)$ be a subsequence of $S(T)$ containing only those statements that are method invocations, and $I_{sc}(T)$ be a subsequence of $I(T)$ containing only those invocations to methods declared in the same class as $T$, and $DM_{Isc}(T)$ be a set containing the distinct methods being invoked by statements from $I_{sc}(T)$, then: $eCouplingTC(T)$ is the number of elements in $DM_{Isc}(T)$.



\subsubsection{$invWithEx(T)$ - Number of method invocations with exceptions}

Different studies have shown that exception-handling code is a neglected area, as is the testing of this code \citep{zhang2012exception,Oliveira16ESEH,Barbosa18GARR,Melo19UEHG}. This is caused by the intrinsic difficulty in these tasks, because it requires special conditions to be recreated just for testing, such as the lack of permissions or hardware problems. But checking for normal and exceptional flow was considered by developers as important (hypothesis H3), therefore we defined two metrics to analyze how much of a test case covers the exception handling possibilities of a method invocation statement.

The first metric determines if a test case is covering a part of the system under test that expects exceptions to be handled, by counting how many of its statements are invocations to methods that have some exception-throwing declaration. It is defined as follows: let $S(T) = S_1, S_2, ..., S_n$ be the sequence of statements contained in test case T's body, and $I(T)$ be a subsequence of $S(T)$ containing only those statements that are method invocations, and $I_{withThrows}(T)$ be a subsequence of $I(T)$ containing only those invocations to methods that have a ``throws'' declaration, then $invWithEx(T)$ is the number of elements in $I_{withThrows}(T)$.

\subsubsection{$invWithExC(T)$ - Number of method invocations with exceptions thrown and caught}

This second metric determines if a test case is indeed providing the expected exception handling code required by the system under test. It is defined as follows: let $Ex(T)$ be a sequence of all exception types being thrown by method invocations in $I_{withThrows}(T)$, and $Ex_{withCatch}(T)$ be a subsequence of $Ex(T)$ containing only those exception types that are being explicitly caught through a ``try-catch'' block or that have a corresponding JUnit's ``expected'' annotation, then $invWithExC(T)$ is the number of elements in $Ex_{withCatch}(T)$.


\subsubsection{$footprint(T)$ - Test case footprint}

This metric corresponds to the amount of lines of code touched by a test case during its execution. This metric excludes the code from the test case itself. It is required for testing hypothesis H14.

\subsection{Data collection}

Data collection was carried out automatically, with the help of different scripts and tools. First, all projects were cloned from GitHub into a local repository.

Regarding the number of LOC, Cyclomatic Complexity and Posnett et al.'s software readability model \citep{posnett2011simpler}, we used scitools's Understand\footnote{\url{https://scitools.com/}} tool, which already provides the basic functions required for calculating these metrics.

For $NumBugFixingCommits(T)$, we used the following procedure: first, we used BEAGLE\footnote{\url{https://github.com/alipourm/testevol2}} to extract detailed commit information from a project's history, including the commit messages informed by the developers.

\begin{sloppypar}
Then, we built a classifier using a dataset provided by \citet{Herbold2022}. The data for training contained annotations made by experts regarding whether the commits are bug fixes or not. The dataset is highly imbalanced, with 449379 commits that were not annotated as bug fixes and 2905 commits that were annotated as bug fixes. We performed a simple downsampling using pandas's\footnote{\url{https://pandas.pydata.org/}} \verb_resample()_ function. Next, we processed the data using NLTK\footnote{\url{https://www.nltk.org/}}, transforming all input into lowercase and removing stopwords based on NLTK's functions and english stopwords list.

Next we splitted the data into train, validation and test subsets, in a proportion of 70\%, 15\% and 15\%, respectively. We also configured two vectorizers: Scikit Learn\footnote{\url{https://scikit-learn.org/stable/}}'s CountVectorizer and TfIdfVectorizer, with parameters obtained with automatic analysis by Auto\_ViML\footnote{\url{https://github.com/AutoViML/Auto_ViML}}. This tool tries different NLP pipelines, and this includes the vectorizers and parameters, which we used for all the experiments.

The experiments were conducted with different algorithms and tools: Auto\_ViML, auto\_sklearn\footnote{\url{https://automl.github.io/auto-sklearn/master/}}, Scikit Learn, XGBoost\footnote{\url{https://xgboost.readthedocs.io/en/stable/}}, LightGBM\footnote{\url{https://lightgbm.readthedocs.io/}} and CatBoost\footnote{\url{https://catboost.ai/}}. These include the following classifiers: LinearSVC, DecisionTreeClassifier, RandomForestClassifier, AdaBoostClassifier, GradientBoostingClassifier,  PassiveAggressiveClassifier, XGBClassifier, LGBMClassifier and CatBoostClassifier. In the end, the best model was LinearSVC with a CountVectorizer, reaching a \textbf{0.97 score} in precision, recall and F1.
\end{sloppypar}


For all the other metrics, we used Eclipse JDT\footnote{\url{https://www.eclipse.org/jdt/}} to perform customized static code analysis. We imported all projects into an Eclipse workspace and made sure all source packages were correctly parsed, without compilation errors. It was not necessary to completely resolve all the dependencies, as JDT can run without them, and we were only interested in the source code of the projects themselves, and not the libraries. JDT has two basic modes: AST (Abstract Syntax Tree), which creates a simple parse tree for the source code, and can be used without a complete Eclipse environment; and Java Model, which includes the AST but also establishes links and connections between Java elements, such as the link between method invocation and method declaration, for example. The Java Model requires the entire Eclipse environment. Our metrics require the Java Model.

For $NumAssertionsWithRecursion(T)$, we had to look into direct and indirect method invocations from a test case in search for assertions. Because this could lead to a possibly infinite recursion and unnecessary diving into hundreds of third-party methods, the scripts were taking an excessively long time to complete. Therefore, we had to limit the number of invocation levels to be inspected. We defined this limit as 5 levels or the re-inspection of a method already inspected before. These limits seemed enough to capture the kind of functionality we expect in a test case in terms of assertions. This difficulty in looking into indirect method invocations was also mentioned by others \citep{delplanque2019rotten}.

For the test case footprint, we used Eclipse's Java Code Coverage tool\footnote{\url{https://www.eclemma.org/}}. For each test case in each project, we generated a bash script that runs that single test case using \verb_mvn test -Dtest=[test case]_. Each script was then individually executed while the code coverage tool monitored the amount of lines of code being covered by that script. The result is the footprint, in terms of lines of code, for each individual test case.

All scripts used to collect and analyze data, the datasets and resulting data sheets are available as open source\footnote{\url{https://github.com/dlucredio/testCaseAnalyzer}}.

\subsection{Dataset}
\label{subsec:datasets}

Our final dataset is composed of 7827 test cases. Most of our metrics consist of discrete values, such as the number of bug-fixing commits in the history of a test case, which is our main measurement of a test case's effectiveness. This may introduce the so-called \textit{discretization noise}. According to \citet{Grano:2021:ieeetse}, this term refers to the introduction of biases due to the presence of data points that are not clearly assignable to a certain class. Therefore, we followed a procedure similar to the one used by \citet{Grano:2021:ieeetse}, dividing the dataset into two sets: \textit{effective} and \textit{ineffective} test cases.

\begin{figure}[htb]
    \centering
    \includegraphics[width=1.0\linewidth]{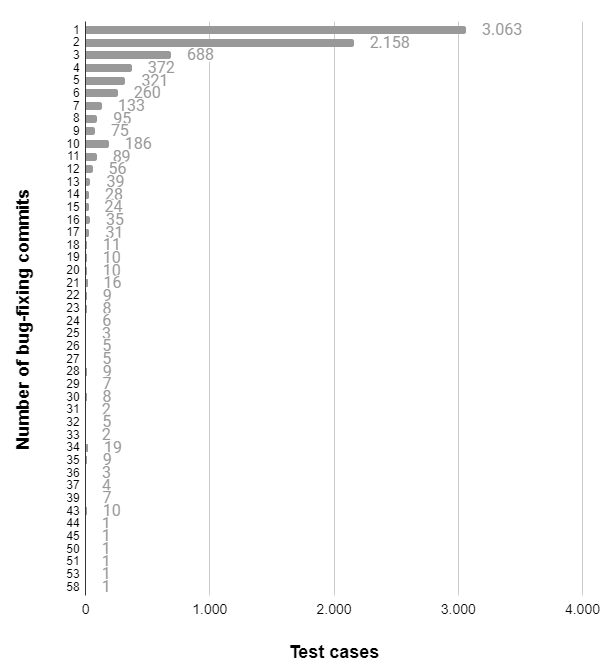}
    \caption{Number of test cases versus number of bug-fixing commits}
    \label{fig:effectiveness}
\end{figure}

\citet{Grano:2021:ieeetse} used quartiles of the mutation score to establish the division between the effective and ineffective sets. In their study, test cases with a mutation score in the first quartile were considered as ineffective and test cases with a mutation score in the fourth quartile were considered as effective. In this way, not only they were able to account for the discretization noise, but they also achieved a good balancing, with 604 effective test cases against 605 ineffective test cases. In our case, this was not possible, because our measurement of effectiveness is a discrete value (number of bug-fixing commits). Also, the distribution was very uneven, with most test cases (3063) having a single bug-fixing commit in its history. Figure \ref{fig:effectiveness} shows the number of test cases versus the number of bug-fixing commits.

Therefore, in order to provide a good balance, instead of quartiles we defined two limits: all test cases with \textbf{one} bug-fixing commit were considered as \textit{ineffective}, and all test cases with \textbf{more than two} bug-fixing commits were considered as \textit{effective}. All 2158 test cases between these limits were discarded as noise. At the end of this process, the \textit{effective} set is composed of 3063 test cases, while the \textit{ineffective} set is composed of 2606 test cases. These limits provided the best balancing possible, as all other combinations resulted in a larger difference of members in each set.




Hypothesis H14 relies on coverage data, as measured by the $footprint$ metric, but there were some practical limitations. Differently from other metrics, which could be collected without actually running the tests, here we had to execute each test. This requires a complete configuration, including all dependencies being correctly resolved. But because this was a manual process, we were not able to do this for all projects. Each one of the projects had a large number of subprojects, and we ended up with more than 400 projects in the Eclipse workspace to manage manually. As a result, we could not obtain observations for all of them, and ended up with 4134 test cases. Although this could be improved with more effort, we consider that this is a sufficiently large number to perform the analysis. The same limits were used here, and as result, the \textit{effective} set (test cases with \textbf{more than two} bug-fixing commits) is composed of 1229 test cases, while the \textit{ineffective}  set (test cases with \textbf{one} bug-fixing commit) is composed of 1285 test cases. Again, this was the best balancing possible. 1620 test cases were discarded as noise. 

Hypothesis H7 states that ``Large test cases are often hard to understand and maintain''. Because this has nothing to do with effectiveness, we used two metrics: $loc(T)$ (number of lines of code in test case) and $cycl(T)$ (cyclomatic complexity). Here, again, the unbalancing is a problem, since most test cases (7247) have $cycl(T)=1$. Because it would be impossible to simply select a smaller value than 1, we did not use groups, as the unbalancing would be too large, and proceeded to use a non-parametric test for correlation analysis using a single group.



\subsection{Statistical analysis}
\label{subsec:analysis}

With these datasets built, we compared the distribution of metrics for each hypothesis in the two sets (effective versus ineffective, in this order) applying the Wilcoxon Rank Sum statistical test with $\alpha$-value = 0.05 as significance threshold. We also estimated the \textit{magnitude} of the observed differences using Cliff's Delta (or $d$), a non-parametric effect size measure for ordinal data. Again, this procedure is the same as the one followed by \citet{Grano:2021:ieeetse}.

The interpretation of the effect size followed well-known guidelines ($|d|<0.147$ is ``negligible'', $|d|<0.33$ is ``small'', $|d|<0.474$ is ``medium'', otherwise ``large''). According to the documentation for the function we used\footnote{\url{https://rdrr.io/cran/rcompanion/man/cliffDelta.html}}, ``When the data in the first group are greater than in the second group, Cliff's delta is positive. When the data in the second group are greater than in the first group, Cliff's delta is negative''. In our case, the first group is the one with effective test cases and the second is the one with ineffective test cases.

For H7, instead of the Wilcoxon Rank Sum test, we performed the Kendall correlation analysis, as the data can not be classified in two groups. It is a non-parametric test, suited for ordinal and not normally distributed data.

\section{Results}
\label{sec:result}

Table \ref{tab:resultsSummary} summarizes the results for each hypothesis. The last column indicates if the data indicates that the hypothesis holds or not. As it can be seen, most hypotheses could not be supported by the data, either because the effect is negligible, the p-value is too large or because the observed effect is the opposite of what was being expected. The exception is H7.

\begin{table*}[htb]
    \caption{Results for the statistical analysis. }
    \label{tab:resultsSummary}
    \centering
    \begin{tabular}{|c|c|c|c|c|c|c|c|} \hline
\textbf{H}	&	\textbf{Method}	&	\textbf{\#}	&	\textbf{Statistic}	&	\textbf{p-value}	&	\textbf{Delta}	&	\textbf{Magnitude}	&	\textbf{Holds?}	\\ \hline
H1	&	Wilcoxon	&	5669	&	4270870.5	&	1.7323e-06	&	0.0701	&	negligible	&	No	\\ \hline
H2a	&	Wilcoxon	&	5669	&	5442881	&	3.4816e-125	&	0.3637	&	medium	&	No	\\ \hline
H2b	&	Wilcoxon	&	5669	&	4412356.5	&	2.5553e-17	&	0.1055	&	negligible	&	No	\\ \hline
H3a	&	Wilcoxon	&	5669	&	3934441.5	&	0.91409	&	-0.0141	&	negligible	&	No	\\ \hline
H3b	&	Wilcoxon	&	5669	&	3900195.5	&	0.99999	&	-0.0227	&	negligible	&	No	\\ \hline
H6	&	Wilcoxon	&	5669	&	4469165.5	&	3.2651e-15	&	0.1197	&	negligible	&	No	\\ \hline
H7	&	Kendall	&	7827	&	0.1143540	&	2.3673e-34	&	N/A	&	small	&	Yes	\\ \hline
H10	&	Wilcoxon	&	5669	&	3915908.5	&	0.9965	&	-0.0188	&	negligible	&	No	\\ \hline
H14	&	Wilcoxon	&	2514	&	953153.5	&	1.2568e-19	&	0.2070	&	small	&	No	\\ \hline
H18	&	Wilcoxon	&	5669	&	2479301	&	1	&	-0.3787	&	medium	&	No	\\ \hline
    \end{tabular}
\end{table*}

Next we discuss the results for each hypothesis.

\subsection{H1: A good test case is specific or atomic, i.e., one test case should be testing one aspect of a requirement}

In the previous survey, this hypothesis received an average score of 3.93 (ranging from 1 and 5). This was not one of the most agreed upon hypotheses, but was considered by many as an important feature for a test case \citep{pavneet2019icse}.

For this hypotheses, we analyzed metrics $assert$ (number of assertions in a test case) against $bfCommits$ (number of bug-fixing commits in the history of a test case). The results for the Wilcoxon Rank Sum have shown a negligible effect size. This means that the number of assertions does not affect the effectiveness of a test case. \textbf{Therefore, we cannot conclude that the hypothesis holds}. 

\subsection{H2: Test cases in a test suite should be self-contained, i.e., independent of one another}	
In the previous survey, this hypothesis received an average score of 3.95 (ranging from 1 to 5), and was also considered as an important feature, mainly to facilitate understanding \citep{pavneet2019icse}.

We analyzed this hypothesis using two pairs of metrics:

\begin{itemize}
    \item H2a. $eCoupling$ (efferent coupling at method level) and $bfCommits$ (number of bug-fixing commits in the history of a test case, in terms of effective versus ineffective test cases); and
    \item H2b. $eCouplingTC$ (efferent coupling for a test case) and $bfCommits$ (number of bug-fixing commits in the history of a test case, in terms of effective versus ineffective test cases).
\end{itemize}

The results for the Wilcoxon Rank Sum have shown a negligible effect for H2b, but a medium positive effect with a small p-value for H2a. A positive delta, in this case, indicates that \textit{effective} test cases have higher coupling values than \textit{ineffective} test cases. This was the opposite to what is stated in the hypothesis, \textbf{therefore we cannot conclude that the hypothesis holds}.

\subsection{H3: Good test cases should check for normal and exceptional flow}

In the previous study, this was one of the most important characteristics of a good test case, with an average score of 4.47 (ranging from 1 to 5). As in the previous hypothesis, here we also used two pairs of metrics:

\begin{itemize}
    \item H3a. $invWithEx$ (number of method invocations with exceptions) and $bfCommits$ (number of bug-fixing commits in the history of a test case, in terms of effective versus ineffective test cases); and
    \item H3b. $invWithExC$ (number of method invocations with exceptions thrown and caught) and $bfCommits$ (number of bug-fixing commits in the history of a test case, in terms of effective versus ineffective test cases).
\end{itemize}

The results for the Wilcoxon Rank Sum have shown a negligible effect size for both tests. This means that the existence of throw/try/catch commands is not associated with the number of bugs detected by a test case. \textbf{Therefore we cannot conclude that the hypothesis holds}. 

\subsection{H6: Most test cases should be small in size (in terms of its lines of code)}

In our previous study this was also not among the most agreed upon hypothesis, with an average score of 3.85 (ranging from 1 to 5), but overall was considered as desirable, as less code is supposed to be better for understanding and simplicity.

For this hypothesis, we analyzed the correlation between $loc$ (number of lines of code in test case) and $bfCommits$ (number of bug-fixing commits in the history of a test case, in terms of effective versus ineffective test cases). A negligible effect was observed for this analysis, \textbf{therefore we cannot conclude that the hypothesis holds}.

\subsection{H7: Large test cases are often hard to understand and maintain}

Among the hypotheses we tested, this hypothesis was the least agreed upon, in our previous survey, with an average score of 3.73 (ranging from 1 to 5).

We analyzed the correlation between $loc$ (number of lines of code in test case) and $cycl$ (McCabe's cyclomatic complexity of test case). All observations were considered (7827). The observed correlation was small but positive, which indicates that, the larger the test case, the more complex it is, thus confirming the hypothesis. Despite the fact that cyclomatic complexity is only an indirect measurement of complexity and difficulty to understand and maintain, \textbf{we can still conclude that the hypothesis holds}. However, the correlation is weak.

\subsection{H10:  Increased complexity in a test case can lead to bugs in the test code itself}

This hypothesis received an average score of 4.04 (ranging from 1 to 5), and was considered as true by most practitioners in our previous survey \cite{pavneet2019icse}.

To test this hypothesis we compared metrics $cycl$ (McCabe's cyclomatic complexity of test case) and $bfCommits$ (number of bug-fixing commits in the history of a test case, in terms of effective versus ineffective test cases). The results for the Wilcoxon Rank Sum have shown a negligible effect size, \textbf{therefore we cannot conclude that the hypothesis holds}. 

\subsection{H14: Each test case should have a small footprint, i.e., the amount of code it executes}

In our previous survey, this was the most controversial hypothesis. Although the average score was 3.92 (ranging from 1 to 5), there were respondents both agreeing and disagreeing with it \citep{pavneet2019icse}.

To test this hypothesis, we used metrics $footprint$ (test case footprint) and $bfCommits$ (number of bug-fixing commits in the history of a test case, in terms of effective versus ineffective test cases). The observed effect was small and positive, which means that \textit{effective} test cases have more footprint than \textit{innefective} test cases, which is the opposite to what is stated in the hypothesis, \textbf{therefore we cannot conclude that the hypothesis holds}.

\subsection{H18: A good test case should be readable and understandable}

This last hypothesis was mostly agreed upon by the practitioners in our previous survey, with an average score of 4.20 (ranging from 1 to 5). We used metrics $read$ (\citet{posnett2011simpler} software readability model for test case) and $bfCommits$ (number of bug-fixing commits in the history of a test case, in terms of effective versus ineffective test cases). 

A medium negative effect was observed, which means that \textit{effective} test cases are less readable/understandable than \textit{ineffective} test cases. This is the opposite of what is stated in the hypothesis, \textbf{therefore we cannot conclude that the hypothesis holds}.


\section{Implications}
\label{sec:implications}

In general, our method failed to obtain supporting evidence for almost all hypotheses from the previous study, with one exception: hypothesis H7 has some evidence supporting it, although the results are not very surprising, since it is expected that cyclomatic complexity increases with LOC. But this is a very intuitive notion that was confirmed both by the previous survey and by this one.

These implications affect software engineering researchers, educators and software engineers. We describe how our findings apply to them and provide opportunities for future work.

\subsection{Researchers}
\label{subsec:researchers}

The most important implication of this study is that some common beliefs may not produce the desired results, as they are only subjective opinions based on experience. We expected to observe at least one confirmation for the hypotheses, but failed to obtain conclusive data, despite our best efforts in obtaining a large number of projects and performing a careful and systematic analysis. Therefore, researchers must be careful when analyzing the results from surveys, always trying to find further evidence to support their underlying truths.


Other researchers are welcome to try different paths than the ones we took here. The first and more obvious one is to replicate the experiment with a different set of projects. Although we focused on good quality projects, perhaps a different set could lead to different observations.

Although we tried different metrics to capture the essence of the investigated hypotheses, there is room for different approaches to be experimented. Our main measurement for most hypotheses was effectiveness (number of bugs found by a test case). This captures the main function of a test case, but it is not the only one. Other quality metrics, for example, the number of test smells \citep{Palomba:2018:icsme} or test case diversity \citep{Feldt:2015:icst}, could be tested in order to validate the hypotheses, as well as refining/modifying the ones we tested here.

Metrics could also be collected differently. Most were obtained from static analysis (number of assertions, coupling, exception handling, lines of code, cyclomatic complexity and readability), but two of them observed different things: test effectiveness was measured by counting commits classified as bug fixes, and test footprint was measured by executing the test cases. Perhaps other ways to collect data could lead to better insight regarding the essence of the hypotheses. Also, we tried to do everything automatically in order to increase the amount of data and to reduce human intervention, but a careful manual approach might provide more comprehensive insight, specially if coupled with machine learning techniques.

\subsection{Educators}
\label{subsec:educators}

Our findings should influence educators on highlighting the difference between belief and evidence regarding good software engineering practices. Although most beliefs represent good advice for achieving quality, students should be taught to gather additional data to determine if these practices are producing the desired results. Many tools and frameworks, such as SonarQube\footnote{\url{https://www.sonarsource.com/products/sonarqube/}}, Qodana\footnote{\url{https://www.jetbrains.com/qodana/}} or CodeScene\footnote{\url{https://codescene.com/}} can be used for this purpose. 

On the other hand, numbers and reports produced automatically should not be trusted blindly. Instead, students must be stimulated to use metrics wisely and always interpret the results in terms of more general software engineering concepts. To cite an example from our study, inserting a lot of try/catch clauses in the code does not necessarily mean that test cases are finding more bugs. Perhaps a good combination of if/else/switch statements could have more impact on quality. The important is to follow the principles behind testing, such as experimenting with boundary values and attempting to find a good sample.

\subsection{Software Engineers}
\label{subsec:softwareEngineers}

The observations made in this study seem to point to a mismatch between the belief of software practitioners and the actual evidence that can be found in software repositories. As a result, developers should understand that some of their beliefs might not lead to the results their expect. For example, separating a large test case into several smaller ones might not lead to more bugs being found, as we observed here.

This also does not mean that their beliefs are false and should just be ignored. We analyzed mainly the effectiveness of a test case in terms of bugs found. Software engineers should not forget that there are other goals to be pursued. Reuse and maintainability, for example, could derive from following these beliefs, even if they do not necessarily lead to better test effectiveness.

Another takeaway from this study is the fact that metrics only tell part of a story. Surely, managers would love to have a large number of reports and summaries that tell them everything that is going right or wrong in their projects and teams. Developers are also particularly happy when they see, for example, a 100\% coverage report popping out after integration. This study should remind software engineers that to consider these numeric signs as a quality certificate is a mistake to be avoided. It is important to conduct serious quality assurance tasks based on solid principles that might use metrics, but as a supporting element, and not the only one.



\section{Threats to Validity}
\label{sec:threats}
In this section, we describe threats to validity for our study. 

\textbf{Construct Validity:} The set of metrics used to evaluate the test cases were selected from the literature, such as LOC, McCabe’s Cyclomatic complexity and Software Readability. However, we had to adapt some source code metrics for the test cases context (efferent coupling for a test case and number of assertions). Before automating the set of metrics, we discussed them with some industrial test engineers who agreed with their rationale and automation. Also, some metrics might have failed to capture the essence of each hypotheses. We mitigated this threat by carefully describing the rationale behind each metric, in order to make it clear how they are related to each hypothesis.

\textbf{External Validity:} We have studied a limited number of Java projects (42 in total, with more than 400 subprojects) and our results cannot be generalized to all projects and domains. However, the selected projects have different sizes and domains and most of them are sponsored by large organizations, such as Google, Microsoft, and Netflix. 

\textbf{Internal Validity:} The primary threats to the internal validity of this study are possible faults in the implementation of our approach and in the tools that we used to perform the evaluation. We control this threat by using different tools validated by other researchers for collecting various metrics (Understand, BEAGLE) and used in industry (JArchitect). We also extensively test our scripts and verify their results against a smaller set of test cases for which we can manually determine the correct results.

\section{Conclusions}
\label{sec:conclusion}

We conducted an extensive research, investigating a large number of software repositories from different large companies. A total of 42 projects, with more than 400 subprojects, were analyzed. We tried to automatically extract as much information as possible to confirm eight practitioners' beliefs related to software testing as surveyed in a previous study \citep{pavneet2019icse}. The data extraction process was not trivial, and although we used some existing tools, many scripts had to be developed for this research.

The results point to a weak confirmation of only one hypothesis from the previous study, while the others were not supported by our analysis. The observations come from data present in good software repositories. It also comes from actual historic data, and not an estimate such as mutation score. Therefore, we can conclude that, as suggested by \citet{devanbu2016belief}, some beliefs expressed by developers might not produce the expected results. 


Our results also indicate possible directions for future work, in order to further investigate the opposite correlations observed for some of the hypotheses. And there are still other hypotheses to test, as many of them were not evaluated here. Some of these required metrics that were considered as too hard or impossible to obtain automatically, therefore another approach is needed. But some were not evaluated due to practical limitations and time constraints, and could be addressed in the near future.

\section*{Acknowledgments}
The authors would like to thank those individuals and institutions who have supported their work on this article.

\bibliographystyle{IEEEtranN}
\bibliography{IEEEabrv,refs}

\begin{thebibliography}{47}
\providecommand{\natexlab}[1]{#1}
\providecommand{\url}[1]{#1}
\csname url@samestyle\endcsname
\providecommand{\newblock}{\relax}
\providecommand{\bibinfo}[2]{#2}
\providecommand{\BIBentrySTDinterwordspacing}{\spaceskip=0pt\relax}
\providecommand{\BIBentryALTinterwordstretchfactor}{4}
\providecommand{\BIBentryALTinterwordspacing}{\spaceskip=\fontdimen2\font plus
\BIBentryALTinterwordstretchfactor\fontdimen3\font minus
  \fontdimen4\font\relax}
\providecommand{\BIBforeignlanguage}[2]{{%
\expandafter\ifx\csname l@#1\endcsname\relax
\typeout{** WARNING: IEEEtranN.bst: No hyphenation pattern has been}%
\typeout{** loaded for the language `#1'. Using the pattern for}%
\typeout{** the default language instead.}%
\else
\language=\csname l@#1\endcsname
\fi
#2}}
\providecommand{\BIBdecl}{\relax}
\BIBdecl

\bibitem[Tassey(2002)]{Tassey2002}
G.~Tassey, ``The economic impacts of inadequate infrastructure for software
  testing,'' 2002.

\bibitem[Kumar and Mishra(2016)]{kunar:procedia:2016}
\BIBentryALTinterwordspacing
D.~Kumar and K.~Mishra, ``The impacts of test automation on software's cost,
  quality and time to market,'' \emph{Procedia Computer Science}, vol.~79, pp.
  8--15, 2016, proceedings of International Conference on Communication,
  Computing and Virtualization (ICCCV) 2016. [Online]. Available:
  \url{https://www.sciencedirect.com/science/article/pii/S1877050916001277}
\BIBentrySTDinterwordspacing

\bibitem[Habib and Pradel(2018)]{Habib:2018:ase}
\BIBentryALTinterwordspacing
A.~Habib and M.~Pradel, ``How many of all bugs do we find? a study of static
  bug detectors,'' in \emph{Proceedings of the 33rd ACM/IEEE International
  Conference on Automated Software Engineering}, ser. ASE 2018.\hskip 1em plus
  0.5em minus 0.4em\relax New York, NY, USA: Association for Computing
  Machinery, 2018, p. 317–328. [Online]. Available:
  \url{https://doi.org/10.1145/3238147.3238213}
\BIBentrySTDinterwordspacing

\bibitem[Adlemo et~al.(2018)Adlemo, Tan, and Tarasov]{Adlemo:2018:wret}
\BIBentryALTinterwordspacing
A.~Adlemo, H.~Tan, and V.~Tarasov, ``Test case quality as perceived in
  sweden,'' in \emph{Proceedings of the 5th International Workshop on
  Requirements Engineering and Testing}, ser. RET '18.\hskip 1em plus 0.5em
  minus 0.4em\relax New York, NY, USA: Association for Computing Machinery,
  2018, p. 9–12. [Online]. Available:
  \url{https://doi.org/10.1145/3195538.3195541}
\BIBentrySTDinterwordspacing

\bibitem[Aniche et~al.(2021)Aniche, Treude, and Zaidman]{Aniche:2021:arxiv}
M.~Aniche, C.~Treude, and A.~Zaidman, ``How developers engineer test cases: An
  observational study,'' 2021.

\bibitem[Grano(2019)]{Grano:2019:issta}
\BIBentryALTinterwordspacing
G.~Grano, ``A new dimension of test quality: Assessing and generating higher
  quality unit test cases,'' in \emph{Proceedings of the 28th ACM SIGSOFT
  International Symposium on Software Testing and Analysis}, ser. ISSTA
  2019.\hskip 1em plus 0.5em minus 0.4em\relax New York, NY, USA: Association
  for Computing Machinery, 2019, p. 419–423. [Online]. Available:
  \url{https://doi.org/10.1145/3293882.3338984}
\BIBentrySTDinterwordspacing

\bibitem[Bavota et~al.(2012)Bavota, Qusef, Oliveto, De~Lucia, and
  Binkley]{bavota:2012:icsm}
G.~Bavota, A.~Qusef, R.~Oliveto, A.~De~Lucia, and D.~Binkley, ``An empirical
  analysis of the distribution of unit test smells and their impact on software
  maintenance,'' in \emph{2012 28th IEEE International Conference on Software
  Maintenance (ICSM)}, 2012, pp. 56--65.

\bibitem[Bavota et~al.(2015)Bavota, Qusef, Oliveto, Lucia, and
  Binkley]{bavota:2014:ese}
\BIBentryALTinterwordspacing
G.~Bavota, A.~Qusef, R.~Oliveto, A.~Lucia, and D.~Binkley, ``Are test smells
  really harmful? an empirical study,'' \emph{Empirical Softw. Engg.}, vol.~20,
  no.~4, p. 1052–1094, Aug. 2015. [Online]. Available:
  \url{https://doi.org/10.1007/s10664-014-9313-0}
\BIBentrySTDinterwordspacing

\bibitem[Bowes et~al.(2017)Bowes, Hall, Petri\'{c}, Shippey, and
  Turhan]{Bowes:2017:wetsm}
D.~Bowes, T.~Hall, J.~Petri\'{c}, T.~Shippey, and B.~Turhan, ``How good are my
  tests?'' in \emph{Proceedings of the 8th Workshop on Emerging Trends in
  Software Metrics}, ser. WETSoM '17.\hskip 1em plus 0.5em minus 0.4em\relax
  IEEE Press, 2017, p. 9–14.

\bibitem[Tosun et~al.(2018)Tosun, Ahmed, Turhan, and Juristo]{Tosun:2018:icssp}
\BIBentryALTinterwordspacing
A.~Tosun, M.~Ahmed, B.~Turhan, and N.~Juristo, ``On the effectiveness of unit
  tests in test-driven development,'' in \emph{Proceedings of the 2018
  International Conference on Software and System Process}, ser. ICSSP
  '18.\hskip 1em plus 0.5em minus 0.4em\relax New York, NY, USA: Association
  for Computing Machinery, 2018, p. 113–122. [Online]. Available:
  \url{https://doi.org/10.1145/3202710.3203153}
\BIBentrySTDinterwordspacing

\bibitem[Bruntink and {van Deursen}(2006)]{BRUNTINK20061219}
\BIBentryALTinterwordspacing
M.~Bruntink and A.~{van Deursen}, ``An empirical study into class
  testability,'' \emph{Journal of Systems and Software}, vol.~79, no.~9, pp.
  1219--1232, 2006, selected papers from the fourth Source Code Analysis and
  Manipulation (SCAM 2004) Workshop. [Online]. Available:
  \url{https://www.sciencedirect.com/science/article/pii/S0164121206000586}
\BIBentrySTDinterwordspacing

\bibitem[Edwards and Shams(2014)]{Edwards:2014:icse}
\BIBentryALTinterwordspacing
S.~H. Edwards and Z.~Shams, ``Comparing test quality measures for assessing
  student-written tests,'' in \emph{Companion Proceedings of the 36th
  International Conference on Software Engineering}, ser. ICSE Companion
  2014.\hskip 1em plus 0.5em minus 0.4em\relax New York, NY, USA: Association
  for Computing Machinery, 2014, p. 354–363. [Online]. Available:
  \url{https://doi.org/10.1145/2591062.2591164}
\BIBentrySTDinterwordspacing

\bibitem[Kochhar et~al.(2019)Kochhar, Xia, and Lo]{pavneet2019icse}
\BIBentryALTinterwordspacing
P.~S. Kochhar, X.~Xia, and D.~Lo, ``Practitioners’ views on good software
  testing practices,'' in \emph{Proceedings of the 41st International
  Conference on Software Engineering: Software Engineering in Practice}, ser.
  ICSE-SEIP ’19.\hskip 1em plus 0.5em minus 0.4em\relax IEEE Press, 2019, p.
  61–70. [Online]. Available:
  \url{https://doi.org/10.1109/ICSE-SEIP.2019.00015}
\BIBentrySTDinterwordspacing

\bibitem[Devanbu et~al.(2016)Devanbu, Zimmermann, and Bird]{devanbu2016belief}
P.~Devanbu, T.~Zimmermann, and C.~Bird, ``Belief \& evidence in empirical
  software engineering,'' in \emph{2016 IEEE/ACM 38th International Conference
  on Software Engineering (ICSE)}.\hskip 1em plus 0.5em minus 0.4em\relax IEEE,
  2016, pp. 108--119.

\bibitem[Grano et~al.(2021)Grano, Palomba, and Gall]{Grano:2021:ieeetse}
G.~Grano, F.~Palomba, and H.~C. Gall, ``Lightweight assessment of test-case
  effectiveness using source-code-quality indicators,'' \emph{IEEE Transactions
  on Software Engineering}, vol.~47, no.~4, pp. 758--774, 2021.

\bibitem[Kracht et~al.(2014)Kracht, Petrovic, and
  Walcott-Justice]{kracht2014empirically}
J.~S. Kracht, J.~Z. Petrovic, and K.~R. Walcott-Justice, ``Empirically
  evaluating the quality of automatically generated and manually written test
  suites,'' in \emph{2014 14th International Conference on Quality
  Software}.\hskip 1em plus 0.5em minus 0.4em\relax IEEE, 2014, pp. 256--265.

\bibitem[Gligoric et~al.(2013)Gligoric, Groce, Zhang, Sharma, Alipour, and
  Marinov]{gligoric2013comparing}
M.~Gligoric, A.~Groce, C.~Zhang, R.~Sharma, M.~A. Alipour, and D.~Marinov,
  ``Comparing non-adequate test suites using coverage criteria,'' in
  \emph{Proceedings of the 2013 International Symposium on Software Testing and
  Analysis}, 2013, pp. 302--313.

\bibitem[Gopinath et~al.(2014)Gopinath, Jensen, and Groce]{gopinath2014code}
R.~Gopinath, C.~Jensen, and A.~Groce, ``Code coverage for suite evaluation by
  developers,'' in \emph{Proceedings of the 36th International Conference on
  Software Engineering}, 2014, pp. 72--82.

\bibitem[Groce et~al.(2014)Groce, Alipour, and Gopinath]{groce2014coverage}
A.~Groce, M.~A. Alipour, and R.~Gopinath, ``Coverage and its discontents,'' in
  \emph{Proceedings of the 2014 ACM International Symposium on New Ideas, New
  Paradigms, and Reflections on Programming \& Software}, 2014, pp. 255--268.

\bibitem[DeMillo et~al.(1978)DeMillo, Lipton, and Sayward]{demillo1978hints}
R.~A. DeMillo, R.~J. Lipton, and F.~G. Sayward, ``Hints on test data selection:
  Help for the practicing programmer,'' \emph{Computer}, vol.~11, no.~4, pp.
  34--41, 1978.

\bibitem[Hamlet(1977)]{hamlet1977testing}
\BIBentryALTinterwordspacing
R.~G. Hamlet, ``Testing programs with the aid of a compiler,'' \emph{IEEE
  Trans. Softw. Eng.}, vol.~3, no.~4, p. 279–290, Jul. 1977. [Online].
  Available: \url{https://doi.org/10.1109/TSE.1977.231145}
\BIBentrySTDinterwordspacing

\bibitem[Jia and Harman(2010)]{jia2010analysis}
Y.~Jia and M.~Harman, ``An analysis and survey of the development of mutation
  testing,'' \emph{IEEE transactions on software engineering}, vol.~37, no.~5,
  pp. 649--678, 2010.

\bibitem[Van~Rompaey et~al.(2007)Van~Rompaey, Du~Bois, Demeyer, and
  Rieger]{van2007detection}
B.~Van~Rompaey, B.~Du~Bois, S.~Demeyer, and M.~Rieger, ``On the detection of
  test smells: A metrics-based approach for general fixture and eager test,''
  \emph{IEEE Transactions on Software Engineering}, vol.~33, no.~12, pp.
  800--817, 2007.

\bibitem[Greiler et~al.(2013)Greiler, Van~Deursen, and
  Storey]{greiler2013automated}
M.~Greiler, A.~Van~Deursen, and M.-A. Storey, ``Automated detection of test
  fixture strategies and smells,'' in \emph{2013 IEEE Sixth International
  Conference on Software Testing, Verification and Validation}.\hskip 1em plus
  0.5em minus 0.4em\relax IEEE, 2013, pp. 322--331.

\bibitem[Palomba and Zaidman(2017)]{palomba2017does}
F.~Palomba and A.~Zaidman, ``Does refactoring of test smells induce fixing
  flaky tests?'' in \emph{2017 IEEE international conference on software
  maintenance and evolution (ICSME)}.\hskip 1em plus 0.5em minus 0.4em\relax
  IEEE, 2017, pp. 1--12.

\bibitem[Begel and Zimmermann(2014)]{begel2014analyze}
A.~Begel and T.~Zimmermann, ``Analyze this! 145 questions for data scientists
  in software engineering,'' in \emph{Proceedings of the 36th International
  Conference on Software Engineering}, 2014, pp. 12--23.

\bibitem[Lo et~al.(2015)Lo, Nagappan, and Zimmermann]{lo2015practitioners}
D.~Lo, N.~Nagappan, and T.~Zimmermann, ``How practitioners perceive the
  relevance of software engineering research,'' in \emph{Proceedings of the
  2015 10th Joint Meeting on Foundations of Software Engineering}, 2015, pp.
  415--425.

\bibitem[Meyer et~al.(2014)Meyer, Fritz, Murphy, and
  Zimmermann]{meyer2014software}
A.~N. Meyer, T.~Fritz, G.~C. Murphy, and T.~Zimmermann, ``Software developers'
  perceptions of productivity,'' in \emph{Proceedings of the 22nd ACM SIGSOFT
  International Symposium on Foundations of Software Engineering}, 2014, pp.
  19--29.

\bibitem[Zou et~al.(2018)Zou, Lo, Chen, Xia, Feng, and
  Xu]{zou2018practitioners}
W.~Zou, D.~Lo, Z.~Chen, X.~Xia, Y.~Feng, and B.~Xu, ``How practitioners
  perceive automated bug report management techniques,'' \emph{IEEE
  Transactions on Software Engineering}, 2018.

\bibitem[Kochhar et~al.(2016)Kochhar, Xia, Lo, and
  Li]{kochhar2016practitioners}
P.~S. Kochhar, X.~Xia, D.~Lo, and S.~Li, ``Practitioners' expectations on
  automated fault localization,'' in \emph{Proceedings of the 25th
  International Symposium on Software Testing and Analysis}, 2016, pp.
  165--176.

\bibitem[Cartaxo et~al.(2016)Cartaxo, Pinto, Vieira, and
  Soares]{cartaxo2016evidence}
B.~Cartaxo, G.~Pinto, E.~Vieira, and S.~Soares, ``Evidence briefings: Towards a
  medium to transfer knowledge from systematic reviews to practitioners,'' in
  \emph{Proceedings of the 10th ACM/IEEE International Symposium on Empirical
  Software Engineering and Measurement}, 2016, pp. 1--10.

\bibitem[Palomba et~al.(2014)Palomba, Bavota, Di~Penta, Oliveto, and
  De~Lucia]{palomba2014they}
F.~Palomba, G.~Bavota, M.~Di~Penta, R.~Oliveto, and A.~De~Lucia, ``Do they
  really smell bad? a study on developers' perception of bad code smells,'' in
  \emph{2014 IEEE International Conference on Software Maintenance and
  Evolution}.\hskip 1em plus 0.5em minus 0.4em\relax IEEE, 2014, pp. 101--110.

\bibitem[Daka and Fraser(2014)]{daka2014survey}
E.~Daka and G.~Fraser, ``A survey on unit testing practices and problems,'' in
  \emph{2014 IEEE 25th International Symposium on Software Reliability
  Engineering}.\hskip 1em plus 0.5em minus 0.4em\relax IEEE, 2014, pp.
  201--211.

\bibitem[Kochhar et~al.(2015)Kochhar, Thung, and Lo]{kochhar2015code}
P.~S. Kochhar, F.~Thung, and D.~Lo, ``Code coverage and test suite
  effectiveness: Empirical study with real bugs in large systems,'' in
  \emph{2015 IEEE 22nd international conference on software analysis,
  evolution, and reengineering (SANER)}.\hskip 1em plus 0.5em minus 0.4em\relax
  IEEE, 2015, pp. 560--564.

\bibitem[Barani et~al.(2023)Barani, Labiche, and Rollet]{Barani2023}
M.~Barani, Y.~Labiche, and A.~Rollet, ``On factors that impact the relationship
  between code coverage and test suite effectiveness: a survey,'' in \emph{2023
  IEEE International Conference on Software Testing, Verification and
  Validation Workshops (ICSTW)}, 2023, pp. 381--388.

\bibitem[Ziftci and Cavalcanti(2020)]{Ziftci2020}
C.~Ziftci and D.~Cavalcanti, ``De-flake your tests : Automatically locating
  root causes of flaky tests in code at google,'' in \emph{2020 IEEE
  International Conference on Software Maintenance and Evolution (ICSME)},
  2020, pp. 736--745.

\bibitem[Rahman and Rigby(2018)]{Rahman2018}
\BIBentryALTinterwordspacing
M.~T. Rahman and P.~C. Rigby, ``The impact of failing, flaky, and high failure
  tests on the number of crash reports associated with firefox builds,'' in
  \emph{Proceedings of the 2018 26th ACM Joint Meeting on European Software
  Engineering Conference and Symposium on the Foundations of Software
  Engineering}, ser. ESEC/FSE 2018.\hskip 1em plus 0.5em minus 0.4em\relax New
  York, NY, USA: Association for Computing Machinery, 2018, p. 857–862.
  [Online]. Available:
  \url{https://doi-org.ez31.periodicos.capes.gov.br/10.1145/3236024.3275529}
\BIBentrySTDinterwordspacing

\bibitem[Posnett et~al.(2011)Posnett, Hindle, and Devanbu]{posnett2011simpler}
D.~Posnett, A.~Hindle, and P.~Devanbu, ``A simpler model of software
  readability,'' in \emph{Proceedings of the 8th working conference on mining
  software repositories}.\hskip 1em plus 0.5em minus 0.4em\relax ACM, 2011, pp.
  73--82.

\bibitem[Xuan and Monperrus(2014)]{xuan2014purification}
\BIBentryALTinterwordspacing
J.~Xuan and M.~Monperrus, ``Test case purification for improving fault
  localization,'' in \emph{Proceedings of the 22nd ACM SIGSOFT International
  Symposium on Foundations of Software Engineering}, ser. FSE 2014.\hskip 1em
  plus 0.5em minus 0.4em\relax New York, NY, USA: Association for Computing
  Machinery, 2014, p. 52–63. [Online]. Available:
  \url{https://doi.org/10.1145/2635868.2635906}
\BIBentrySTDinterwordspacing

\bibitem[Delplanque et~al.(2019)Delplanque, Ducasse, Polito, Black, and
  Etien]{delplanque2019rotten}
\BIBentryALTinterwordspacing
J.~Delplanque, S.~Ducasse, G.~Polito, A.~P. Black, and A.~Etien, ``Rotten green
  tests,'' in \emph{Proceedings of the 41st International Conference on
  Software Engineering}, ser. ICSE ’19.\hskip 1em plus 0.5em minus
  0.4em\relax IEEE Press, 2019, p. 500–511. [Online]. Available:
  \url{https://doi.org/10.1109/ICSE.2019.00062}
\BIBentrySTDinterwordspacing

\bibitem[Zhang and Elbaum(2012)]{zhang2012exception}
P.~Zhang and S.~Elbaum, ``Amplifying tests to validate exception handling
  code,'' in \emph{Proceedings of the 34th International Conference on Software
  Engineering}, ser. ICSE ’12.\hskip 1em plus 0.5em minus 0.4em\relax IEEE
  Press, 2012, p. 595–605.

\bibitem[Oliveira et~al.(2016)Oliveira, Cacho, Borges, Silva, and
  Castor]{Oliveira16ESEH}
\BIBentryALTinterwordspacing
J.~Oliveira, N.~Cacho, D.~Borges, T.~Silva, and F.~Castor, ``An exploratory
  study of exception handling behavior in evolving android and java
  applications,'' in \emph{Proceedings of the 30th Brazilian Symposium on
  Software Engineering}, ser. SBES'16.\hskip 1em plus 0.5em minus 0.4em\relax
  New York, NY, USA: Association for Computing Machinery, 2016, pp. 23--32.
  [Online]. Available: \url{https://doi.org/10.1145/2973839.2973843}
\BIBentrySTDinterwordspacing

\bibitem[Barbosa and Garcia(2018)]{Barbosa18GARR}
E.~A. Barbosa and A.~Garcia, ``Global-aware recommendations for repairing
  violations in exception handling,'' \emph{IEEE Transactions on Software
  Engineering}, vol.~44, no.~9, pp. 855--873, 2018.

\bibitem[Melo et~al.(2019)Melo, Coelho, and Treude]{Melo19UEHG}
H.~Melo, R.~Coelho, and C.~Treude, ``Unveiling exception handling guidelines
  adopted by java developers,'' in \emph{2019 IEEE 26th International
  Conference on Software Analysis, Evolution and Reengineering (SANER)}, 2019,
  pp. 128--139.

\bibitem[Herbold et~al.(2022)Herbold, Trautsch, Ledel, Aghamohammadi, Ghaleb,
  Chahal, Bossenmaier, Nagaria, Makedonski, Ahmadabadi, Szabados, Spieker,
  Madeja, Hoy, Lenarduzzi, Wang, Rodr{\'i}guez-P{\'e}rez, Colomo-Palacios,
  Verdecchia, Singh, Qin, Chakroborti, Davis, Walunj, Wu, Marcilio, Alam,
  Aldaeej, Amit, Turhan, Eismann, Wickert, Malavolta, Sul{\'i}r, Fard, Henley,
  Kourtzanidis, Tuzun, Treude, Shamasbi, Pashchenko, Wyrich, Davis, Serebrenik,
  Albrecht, Aktas, Str{\"u}ber, and Erbel]{Herbold2022}
\BIBentryALTinterwordspacing
S.~Herbold, A.~Trautsch, B.~Ledel, A.~Aghamohammadi, T.~A. Ghaleb, K.~K.
  Chahal, T.~Bossenmaier, B.~Nagaria, P.~Makedonski, M.~N. Ahmadabadi,
  K.~Szabados, H.~Spieker, M.~Madeja, N.~Hoy, V.~Lenarduzzi, S.~Wang,
  G.~Rodr{\'i}guez-P{\'e}rez, R.~Colomo-Palacios, R.~Verdecchia, P.~Singh,
  Y.~Qin, D.~Chakroborti, W.~Davis, V.~Walunj, H.~Wu, D.~Marcilio, O.~Alam,
  A.~Aldaeej, I.~Amit, B.~Turhan, S.~Eismann, A.-K. Wickert, I.~Malavolta,
  M.~Sul{\'i}r, F.~Fard, A.~Z. Henley, S.~Kourtzanidis, E.~Tuzun, C.~Treude,
  S.~M. Shamasbi, I.~Pashchenko, M.~Wyrich, J.~Davis, A.~Serebrenik,
  E.~Albrecht, E.~U. Aktas, D.~Str{\"u}ber, and J.~Erbel, ``A fine-grained data
  set and analysis of tangling in bug fixing commits,'' \emph{Empirical
  Software Engineering}, vol.~27, no.~6, p. 125, Jul 2022. [Online]. Available:
  \url{https://doi.org/10.1007/s10664-021-10083-5}
\BIBentrySTDinterwordspacing

\bibitem[Palomba et~al.(2018)Palomba, Zaidman, and
  De~Lucia]{Palomba:2018:icsme}
F.~Palomba, A.~Zaidman, and A.~De~Lucia, ``Automatic test smell detection using
  information retrieval techniques,'' in \emph{2018 IEEE International
  Conference on Software Maintenance and Evolution (ICSME)}, 2018, pp.
  311--322.

\bibitem[Feldt et~al.(2015)Feldt, Poulding, Clark, and Yoo]{Feldt:2015:icst}
R.~Feldt, S.~Poulding, D.~Clark, and S.~Yoo, ``Test set diameter: Quantifying
  the diversity of sets of test cases,'' in \emph{2016 IEEE International
  Conference on Software Testing, Verification and Validation (ICST)}, 2015,
  pp. 223--233.

\end{thebibliography}

\vfill

\end{document}